\documentclass[preprint,pre,runinaddress]{revtex4}
\usepackage{amsmath}
\usepackage{amssymb}

\input{tcilatex}

\begin{document}

\title{Energy localization in the Peyrard-Bishop DNA model}
\author{Jayme De Luca$^{1}$ }
\email{ deluca@df.ufscar.br}
\author{ Elso Drigo Filho$^{2}$, Antonio Ponno$^{3}$ and Jos\'{e} Roberto
Ruggiero$^{2}$}
\affiliation{$^{1}$Universidade Federal de S\~{a}o Carlos, \\
Departamento de F\'{\i}sica\\
Rodovia Washington Luis, km 235\\
S\~{a}o Carlos, 13565-905-S\~{a}o Paulo, Brazil\\
$^{2}$Universidade Estadual Paulista "Julio de Mesquita Filho", \ \\
Instituto de Bioci\^{e}ncias, letras e Ci\^{e}ncias Exatas, Departamento de F%
\'{\i}sica, Rua Cristov\~{a}o Colombo 2265, \ S\~{a}o Jos\'{e} do Rio Preto,
15054-000-S\~{a}o Paulo, Brazil\\
$^{3}$Universit\`{a} degli Studi di Milano, Dipartimento di Matematica \ "F.
Enriques "-Via Saldini 50, 20133 Milano, Italy}
\date{\today }

\begin{abstract}
We study energy localization on\ the oscillator-chain proposed by Peyrard
and Bishop to model DNA. We search \ numerically for conditions with
initial energy in a small subgroup of consecutive oscillators of a finite
chain and such that the oscillation amplitude is small outside this subgroup
on a long timescale. We use a localization criterion based on the
information entropy and we verify numerically that such localized
excitations exist when the nonlinear dynamics of the subgroup oscillates
with a frequency inside the reactive band of the linear chain. We predict a
mimium value for the Morse parameter $(\mu >2.25)$ (the only parameter of
our normalized model), in agreement with the numerical calculations (an
estimate for the biological value is $\mu =6.3$). For supercritical masses,
we use canonical perturbation theory to expand the frequencies of the
subgroup and we calculate an energy threshold in agreement with the
numerical calculations.
\end{abstract}

\maketitle

\section{Introduction}

\bigskip

A plethora of chemical processes involving the DNA macromolecule are known 
\cite{Saenger,Calladine,PeyrardReview}, for example the existence of
denaturation bubbles containing a few broken H-bonds, and the transcription
process triggered by the bonding of the biochemical complex to a specific
region of the DNA (the so called TATA-box). The oscillator-chain model for
DNA \cite{PB89} was first proposed to study the thermal denaturation of
the DNA macromolecule, i.e. the separation of the two strands. The dynamics
of this model was first approximated with soliton techniques\cite%
{TDP89,secondGaeta,G99,BCPR99}. Our motivation for the present work was to
study this model with methods of finite-dimensional dynamical systems, which
could later be extended to a realistic model of the DNA without translation
symmetry. In this work we consider a finite chain of $N$ oscillators with
initial condition restricted to a small group of $n<<N$ \ of consecutive
oscillators. We define a localized motion as one in which the amplitude of
oscillation is small outside a group of $n_{\max }$ oscillators for all
times, with $n<n_{\max }<<N$ , and we introduce a numerical criterion to
quantify localization based on the information entropy. We use the
correspondence conjecture (CC) of Flach et al. \cite{FWO94,FW98} that the
nonlinear dynamics of the isolated group of $n$ oscillators must have
frequencies inside the reactive band of the linearized chain for
localization to be possible. Within this conjecture, we show that there is a
minimum value for the Morse parameter (the only parameter of the model) for
a localized excitation to be possible. The predicted value $\mu =2.25$
agrees with our numerical calculations. Last, since the linear frequencies
of the isolated $n$-system lie in the dispersive band, an immediate
consequence of the (CC) is that there must be a critical nonzero energy for
localization (namely, for at least one of the $n$ frequencies to exit the
dispersive band). For supercritical values of the Morse parameter, we use
canonical perturbation theory to evaluate the frequency shifts and we
predict a threshold energy for localization in agreement with the numerical
calculations.

For the normalizations that follow, the most convenient way to introduce the
Peyrard-Bishop (BP) model \cite{PB89} for a DNA macromolecule is by the
Lagrangian

\begin{equation}
L_{PB}=\sum_{i=1}^{N}\frac{m}{2}(\dot{u}_{i}^{2}+\dot{v}_{i}^{2})-\frac{k}{2}%
(u_{i+1}-u_{i})^{2}-\frac{k}{2}(v_{i+1}-v_{i})^{2}-D[\exp
a(v_{i}-u_{i})-1]^{2},  \label{PBmodel}
\end{equation}%
where $u_{i}\ $and $v_{j}$ denote the relative displacements of the
nucleotidic bases at sites $i$ and $j$ of each DNA strand, respectively. The
number $N$ denotes the number of sites in each strand of the DNA and can be
as large as $N\sim 10^{9}$. For practical reasons we perform our numerical
experiments with up to $N=500$. The masses of the bases have a common value $%
m$, the constant $k$ is the longitudinal elastic constant and the parameters 
$D$ and $a$ define the Morse potential describing the transversal H-bonds
linking the two chains. The experimental values for these parameters have been
discussed in the literature\cite{ZG76,EZ91,SZ2001,PeyrardReview}: the mass
of the base pairs is about ($300$ $a.m.u.=5.010^{-25}kg$), and the linear
spring constant is ($0.04eV/\mathring{A}^{2}$). The hydrogen bond is
modelled by the Morse potential with $a=4.45\mathring{A}^{-1}$ and for D we
take an average of the value for the guanine-citosine basis (G-C) and the
value for the thymine-adenine, A-T base pair, $\bar{D}=0.04eV$ \cite{ZG76}

\bigskip By means of a rotation of coordinates defined by 
\begin{eqnarray}
x_{i} &=&(u_{i}+v_{i})/\sqrt{2},  \label{rotation} \\
y_{i} &=&(u_{i}-v_{i})/\sqrt{2},  \notag
\end{eqnarray}%
the PB Lagrangian (\ref{PBmodel}) splits into the sum $L_{BP}=L_{X}+L_{Y}$,\
with%
\begin{equation}
L_{X}=\sum_{i=1}^{N}m\dot{x}_{i}^{2}/2-k(x_{i+1}-x_{i})^{2}/2,  \label{LX}
\end{equation}%
\ depending only on the $x$ coordinates and with $L_{Y}$ depending only on
the $y$ coordinates as

\begin{equation}
L_{Y}=\dsum\limits_{i=1}^{N}\frac{1}{2}m\dot{y}_{i}^{2}-\frac{k}{2}%
(y_{i+1}-y_{i})^{2}-\bar{D}[\exp (-\sqrt{2}ay_{i})-1]^{2}.  \label{LY}
\end{equation}%
Lagrangian $L_{Y}$ can be normalized by introducing a dimensionless time
parameter $\tau \equiv \sqrt{k/m}t$ and a dimensionless coordinates $\xi
_{i}\equiv \sqrt{2}ay_{i}$. The above scalings bring $L_{Y}$ of Eq. (\ref{LY}%
) to the normal form $L_{Y}\equiv \frac{k}{2a^{2}}L$, with%
\begin{equation}
L=\dsum_{i=1}^{N}\frac{1}{2}\dot{\xi}_{i}^{2}-\frac{1}{2}(\xi _{i+1}-\xi
_{i})^{2}-\frac{\mu ^{2}}{2}[\exp (-\xi _{i})-1]^{2},  \label{NormalL}
\end{equation}%
where dot denotes derivative with respect to $\tau $. \ Our normalization
differs from that of \cite{PeyrardReview}, and it was chosen such that the
quartic approximation to Eq. (\ref{LY}) has the form of the Klein-Gordon
oscillator chain studied in \cite{DL2002,Monica-Sola}. We henceforth study a
chain of $N$ sites with periodic boundary conditions described by Lagrangian
(\ref{NormalL}), a dynamical system depending on the single parameter $\mu
^{2}\equiv 4\bar{D}a^{2}/k$ (henceforth called the Morse parameter). Using
the values of the literature\cite{ZG76,EZ91,SZ2001}, we estimate a realistic
biological value for $\mu $ to be $\mu =6.3$, and the scaling factors for
the units of time and energy to be $0.810^{-12}s$ and $2.010^{-3}eV$
respectively.

For small amplitudes, the normal mode frequency spectrum of (\ref{NormalL})
is \cite{DL2002}

\begin{equation}
\omega (k)=\sqrt{\mu ^{2}+4\sin ^{2}(k/2)},  \label{dispersion}
\end{equation}%
where $k=(j\pi /N),$ $j=1,...N$. The range of normal mode frequencies $\mu
\leq \omega \leq \sqrt{\mu ^{2}+4}$ constitutes the dispersive band, while
the two relations $0\leq \omega <\mu $ and $\omega >\sqrt{\mu ^{2}+4}$
define the lower and upper reactive bands, respectively. One expects that
localized motions of the chain with frequency components inside the
dispersive band will give rise to quasi-normal modes excitations, which are
typically delocalized in space. In such a way the localized state looses its
energy in the form of radiation and spreads out. In contrast, localized
excitations displaying only frequency components inside the two reactive
bands are expected to preserve localization for long times.

There is a large body of studies of one-dimensional chains, investigating
the energy interchange among the \ \textquotedblleft linearized
\textquotedblright\ system modes \cite%
{Chaos5,PRE1,PRE2,Driscoll,Burlakov,Cretegny,Kosevich,Mirnov}. For initial
energy in a few low frequency modes, one of us (J. DL) has developed
theoretical descriptions for energy spreading among modes, valid in various
energy ranges, which were compared to numerical results for the FPU chain 
\cite{Chaos5,PRE1,PRE2} and for the Klein-Gordon chain \cite{DL2002}. If the
energy is initially placed in high frequency modes, the dynamics is
transiently mediated by the formation of unstable nonlinear structures\cite%
{Burlakov,Cretegny,Kosevich,Mirnov}. The mode energy is found to distribute
itself first into a number of structures, localized in space, each
consisting of a few oscillators, which coalesce over time into a single
localized structure, a chaotic breather (CB). Over longer times the CB is
found to break up, with energy transferred to lower frequency modes.
Recently, there have been another set of studies of the discretized
Klein-Gordon equation, from the perspective of studying the stability of
breathers, which are chosen as initial conditions \cite{Bang,Marin}. For a
more comprehensive discussion of the extensive research on the dynamics of
oscillator chains, see for example Ref. \cite{DL2002}.

The study of \ soliton solutions of the nonlinear PDE's obtained by
multiple-scale expansions constitutes at present the main line of study of
the nonlinear dynamics of the\ DNA models\cite{TDP89,secondGaeta,BCPR99}%
\textbf{. } Even though the use of modulation equations and soliton theory
does furnish interesting results, we made here the choice to follow a
different approach, based on normal form methods for low-dimensional
Hamiltonian dynamical systems. The reason for this choice is that, as is
well known, multiple-scale expansions are only valid for initial data
varying slowly in real space and quasi-monochromatic in Fourier space, while
here we are interested in the evolution of initial excitations strongly
localized in real space (delocalized in Fourier space). Moreover, the method
used here displays the further advantage that it could be suitably extended
to inhomogeneous chain models describing a realistic DNA molecule.

\section{\protect\bigskip The correspondence conjecture revisited}

In what follows we revisit the correspondence conjecture (CC) of Refs. \cite%
{FWO94,FW98} in the light of\ canonical perturbation theory. With reference
to system (\ref{NormalL}), let us initially displace from the equilibrium
position $\mathbf{\xi =0}$ only a finite number $n<<N$ of consecutive
particles. For such initial datum, instead of studying the dynamics of the
full chain, involving a large number of degrees of freedom, we study the
dynamics of the subsystem defined by the Lagrangian (\ref{NormalL}), with the
sum restricted to the sites corresponding to the degrees of freedom
initially excited, and with fixed end boundary conditions for the next
neighbors. Such an $n$-degree of freedom subsystem is thought of as isolated
and having energy $E$. This subsystem can be regarded as a perturbation of $%
n $ linearly coupled oscillators, whose normal mode frequencies $\omega
_{1},\ldots ,\omega _{n}$ are shown to lie inside the dispersive band of the
larger $N$-chain. For sufficiently low energies, the dynamics is
quasi-linear and its frequency spectrum is close to the normal mode
frequencies $\omega _{1},\ldots ,\omega _{n}$. According to the (CC), if one
uses such initial conditions for the larger lattice, the normal modes of the
larger chain are excited and the initial excitation will spread out, which
is what we observe numerically. With increasing energy, the effect of
nonlinearity becomes prominent, and the frequency spectrum is modified. In
the absence of resonances of third and fourth order (at least) in the
harmonic spectrum of the subsystem, the modes preserve their identity and we
can follow their frequency shifts inside the dispersive band. According to
the (CC), one has localization for initial data at a given energy when the
frequencies of the corresponding motion of the subsystem is outside the
dispersive band. Of course this can happen only if the energy (i.e. the
nonlinearity) is high enough.

The conditions required on the frequency spectrum for multi-periodic
oscillations are much more restrictive and the localization properties of
such states can be very weak\cite{FW98}. For this reason, we restrict the
analysis to \textit{periodic }oscillations only, which amounts to look for
periodic orbits of the subsystem whose frequency and harmonics lay outside
the dispersive band. This analysis is detailed below.

\subsection{\protect\bigskip Analysis of the finite subsystems}

For the theoretical analysis we consider localized excitations where the
amplitude of oscillation is small \ for sites outside a group of $1+2r$
modes, say $|i|>r$ (our $i$ runs to both negative and positive directions
and the central particle is $i=0$). Under this conjecture the dynamics for
sites on the right-hand side of the group ($i>r$) \ can be approximated by a
linear chain driven by the given oscillation of oscillator $r$ (while the
same can be said of the oscillators on the left-hand side, $i<-r$). The
equation of motion for the linearized chain can be derived from the 
Lagrangian (\ref{NormalL}) by expanding the exponential
\begin{equation}
\ddot{\xi}_{i}=\xi _{i+1}+\xi _{i-1}-(2+\mu ^{2})\xi _{i}\quad ,\quad i>r
\label{linear}
\end{equation}%
where the above linearization holds only for the oscillators outside the
subgroup, which are supposed to oscillate with a small amplitude ($i>r$).
The coordinate$\ \xi _{r}(t)$ of oscillator $r$ entering into Eq. (\ref%
{linear}) must be given \emph{a priori} as a known forcing term. To solve
Eq.(\ref{linear}) by Fourier transform we define the two-component
vector 
\begin{equation}
\chi _{i+1}\equiv 
\begin{pmatrix}
x_{i+1}(\omega ) \\ 
x_{i}(\omega )%
\end{pmatrix}%
.  \label{defivec}
\end{equation}%
It can be shown with the help of Eq. (\ref{linear}) that $\chi _{i+1}$%
satisfies the linear matrix iteration law

\begin{equation}
\chi _{i+1}=%
\begin{pmatrix}
(\omega _{o}^{2}-\omega ^{2}) & -1 \\ 
1 & 0%
\end{pmatrix}%
\chi _{i},  \label{iterate}
\end{equation}%
where $\omega _{o}^{2}\equiv 2+\mu ^{2}$. For example in the case of a
monochromatic forcing, $\chi _{r+1}(\omega )$ is nonzero only at a single
frequency $\bar{\omega}$, and for the iteration of Eq.(\ref{iterate}) to
produce a bounded amplitude for sites of a large $i$ it is necessary that $(%
\bar{\omega}^{2}-\omega _{o}^{2})^{2}>4$, which is the definition of the
reactive band (as opposed to the radiation band defined by $\mu <\omega <%
\sqrt{\mu ^{2}+4}$). If the forcing has several large Fourier components,
the first large component might be in the lower reactive band ($\omega <\mu $
), while the other important harmonics could be in the upper reactive band ($%
\omega >\sqrt{4+\mu ^{2}}$).

The first subsystem we consider here (henceforth called the 1-system) is
defined by $r=0$ and consists of the nonlinear oscillation of a single
particle of coordinate $x_{o}(t)$ with fixed ends ($\xi _{-1}=\xi _{1}=0$ ).
This nonlinear dynamics can be derived from Lagrangian (\ref{NormalL}) and
it is also described by the Hamiltonian%
\begin{equation}
H=\frac{1}{2}p^{2}+\xi _{o}^{2}+\frac{\mu ^{2}}{2}(\exp (-\xi _{o})-1)^{2}.
\label{hamiltonian1}
\end{equation}%
The frequency of oscillation for the periodic motion of \ Hamiltonian (\ref%
{hamiltonian1}) can be determined by a simple quadrature for any energy, by
the formula%
\begin{equation}
\omega =\pi \left[ \dint\limits_{\xi \min }^{\xi \max }\frac{d\xi _{o}}{%
\sqrt{2E-2\xi _{o}^{2}-\mu ^{2}(\exp (-\xi _{o})-1)^{2}}}\right] ^{-1}.
\label{quadrature}
\end{equation}%
In figure 1 we plot this frequency as a function of the energy for several
values of the parameter $\mu $ to illustrate that it is always inside the
radiation band for $\mu <2.25$ at any energy. This is then the minimum value
for the $\mu $ parameter where localization is possible, as predicted by the
correspondence conjecture for the simple 1-system. It turns out that the
biological value is $\mu =6.3>2.25$ , in agreement with this theory. Another
agreement with this simple theory is discussed in the numerical section, as
the numerical searches never found a localized state with $\mu <2.5$.

For supercritical values of $\mu $ ($\mu >2.25$ ), the frequency (\ref%
{quadrature}) is in the lower reactive band for a sufficient large energy.
The frequency of small oscillations (zero energy) is easily obtained by
expanding Hamiltonian (\ref{hamiltonian1}) to quadratic order, and is $%
\omega _{o}=\sqrt{2+\mu ^{2}}>\mu .$ The next correction for small energies
can be obtained by expanding Hamiltonian (\ref{hamiltonian1}) to fourth
order in $\xi _{o}$ as%
\begin{equation}
H_{1}^{(4)}=\frac{1}{2}p^{2}+\frac{1}{2}\omega _{o}^{2}\xi _{o}^{2}-\frac{%
\mu ^{2}}{2}\xi _{o}^{3}+\frac{7\mu ^{2}}{24}\xi _{o}^{4},  \label{hami1}
\end{equation}%
where the superscript and subscript on $H$ refer to the order of the
expansion and to the 1-system, respectively. Introducing action-angle
variables and using standard canonical perturbation theory \cite{Galgani},
we find that the normal form of Hamiltonian (\ref{hami1}) up to second order
in the action variable $J$ is 
\begin{equation}
\hat{H}_{1}^{(4)}=\omega _{o}J-g(\mu )J^{2},  \label{canonical1}
\end{equation}%
with 
\begin{equation}
g(\mu )=\frac{(4\mu ^{2}-7)}{8(\mu ^{2}+2)^{2}}.  \label{defg}
\end{equation}%
Notice that for supercritical values of $\mu $ ($\mu >2.25$), the
coefficient $g(\mu )$ as defined by Eq. (\ref{defg}) is positive ($g(\mu )>0$%
), such that the nonlinear frequency decreases with increasing energy.
Defining the nonlinear frequency by $\Omega \equiv \partial
H_{1}^{(4)}/\partial J$, the conditions $H_{1}^{(4)}=E_{c}$ and $\Omega
=\partial H_{1}^{(4)}/\partial J=\mu $ determine the minimum energy $E_{c}$
to be

\begin{equation}
E_{c}=\frac{\omega _{o}^{2}-\mu ^{2}}{4g(\mu )}=\frac{4(\mu ^{2}+2)^{2}}{\mu
^{2}(4\mu ^{2}-7)}.  \label{Ecritical1}
\end{equation}%
The interpretation of Eq. (\ref{Ecritical1}) is as follows: If the isolated
nonlinear oscillator of the 1-system has an energy $E>E_{c}$, its frequency
is in the reactive band $(\Omega <\mu )$ and we expect that the
corresponding type (i) initial condition should produce a localized
excitation, according to (CC). This determination of the critical energy is
compared to the numerical results in the following section, and it turns out
to be short by a factor of two. The explanation for this is that the
subsystem consisting of a single oscillator looses a significant amount of
energy to the immediate neighbors, such that one could expect a higher
critical energy. It turns out that the value $E_{c}\simeq 1$ predicted by
Eq.(\ref{Ecritical1}) \ is precisely a factor of two short of the numerical
value $E_{c}\simeq 2$ for any value of $\mu $. Our simple theory is then
seen to be only in qualitative agreement with the numerical calculations. A
better approximation should be given by a subsystem consisting of three
particles with fixed ends, which is our next subsystem.

We consider another subsystem (henceforth called the 3-system), consisting
of three oscillators along the symmetric motion defined by $\xi _{-1}=\xi
_{1}$. The Lagrangian equations of motion derived from (\ref{NormalL}) with
the condition $\xi _{-1}=\xi _{1}$ correspond to the following two-degree of
freedom Hamiltonian 
\begin{eqnarray}
H_{3} &=&\frac{1}{2}p_{o}^{2}+\frac{1}{4}p_{1}^{2}+\xi _{1}^{2}+(\xi
_{o}-\xi _{1})^{2}  \label{hamiS} \\
&&+\mu ^{2}(\exp (-\xi _{1})-1)^{2}+\frac{\mu ^{2}}{2}(\exp (-\xi
_{o})-1)^{2}.  \notag
\end{eqnarray}%
The two frequencies of the quasi-periodic linear motion at zero energy are 
\begin{eqnarray}
\omega _{1} &=&\ \sqrt{\mu ^{2}+2-\sqrt{2}},  \label{linear3freq} \\
\omega _{2} &=&\sqrt{\mu ^{2}+2+\sqrt{2}},  \notag
\end{eqnarray}%
\ \ which are inside the dispersive band for any $\mu $. \ 

For small energies subsystem (\ref{hamiS}) is a perturbation of \ two
harmonic oscillators with frequencies $\omega _{1}$and $\omega _{2}$ inside
the dispersive band of the whole linearized chain. To compute the leading
contribution to the frequency shift of each oscillator we must evaluate the
next frequency correction in powers of the actions. One can check that there
is no resonance up to fourth order involving the linear part of (\ref{hamiS}%
), i.e. $\omega _{1}/\omega _{2}\neq 1/2,1/3$. Using canonical perturbation
theory\ \cite{Galgani} we can remove the cubic term from (\ref{hamiS}),
average the quartic term and express the normal form of Hamiltonian (\ref%
{hamiS}) up to second order in the actions as 
\begin{equation}
\hat{H}_{3}^{(4)}(J_{1},J_{2})=\omega _{1}J_{1}+\omega
_{2}J_{2}-c_{1}J_{1}^{2}-c_{2}J_{2}^{2}-c_{12}J_{1}J_{2},  \label{Hamsubn}
\end{equation}%
where the $J$'s are the action variables and $c_{1}$, $c_{2}$ and $c_{12}$
are given by%
\begin{eqnarray}
c_{1} &=&\frac{3\mu ^{2}\left[ 12\mu ^{6}+(65-26\sqrt{2})\mu ^{4}+(73-75%
\sqrt{2})\mu ^{2}-(42-35\sqrt{2})\right] }{64\omega _{1}^{4}\omega
_{2}^{2}\left( 4\omega _{1}^{2}-\omega _{2}^{2}\right) },  \label{c1c2c12} \\
c_{12} &=&\frac{3\mu ^{2}\left[ 36\mu ^{8}+147\mu ^{6}+4\mu ^{4}-278\mu
^{2}-98\right] }{16\omega _{1}^{3}\omega _{2}^{3}\left( 4\omega
_{1}^{2}-\omega _{2}^{2}\right) \left( 4\omega _{2}^{2}-\omega
_{1}^{2}\right) },  \notag \\
c_{2} &=&\frac{3\mu ^{2}\left[ 12\mu ^{6}+(65+26\sqrt{2})\mu ^{4}+(73+75%
\sqrt{2})\mu ^{2}-(42+35\sqrt{2})\right] }{64\omega _{1}^{2}\omega
_{2}^{4}\left( 4\omega _{2}^{2}-\omega _{1}^{2}\right) }.  \notag
\end{eqnarray}%
The nonlinearly modified frequencies are given by

\begin{eqnarray}
\Omega _{1} &=&\frac{\partial \hat{H}}{\partial J_{1}}=\omega
_{1}-2c_{1}J_{1}-c_{12}J_{2},  \label{nonlinearfreqs} \\
\Omega _{2} &=&\frac{\partial \hat{H}}{\partial J_{2}}=\omega
_{2}-2c_{2}J_{2}-c_{12}J_{1}.  \notag
\end{eqnarray}

For supercritical values of $\mu $ the coefficients of (\ref{c1c2c12}) are
all positive, such that the frequencies of (\ref{nonlinearfreqs}) are
decreasing functions of the energy. The two periodic orbits branching from
the linear modes of (\ref{Hamsubn}) are obtained by setting one of the
actions\ of (\ref{Hamsubn}) to zero. For example by substituting $J_{2}=0$
into (\ref{Hamsubn}) we obtain 
\begin{equation}
\hat{H}_{3}^{(4)}(J_{1},0)=\omega _{1}J_{1}-c_{1}J_{1}^{2},  \label{type1}
\end{equation}%
such that the critical energy predicted by setting $\Omega _{1}=\mu $ is%
\begin{equation}
E_{c}^{(1)}=\frac{\omega _{1}^{2}-\mu ^{2}}{4c_{1}}=\frac{2-\sqrt{2}}{4c_{1}}%
.  \label{energy1}
\end{equation}%
For the other periodic orbit we substitute $J_{1}=0$ into (\ref{Hamsubn}),
yielding 
\begin{equation}
\hat{H}_{3}^{(4)}(0,J_{2})=\omega _{2}J_{2}-c_{2}J_{2}^{2},  \label{type2}
\end{equation}%
and the critical energy predicted by setting $\Omega _{2}=\mu $ is%
\begin{equation}
E_{c}^{(2)}=\frac{\omega _{2}^{2}-\mu ^{2}}{4c_{2}}=\frac{2+\sqrt{2}}{4c_{2}}%
.  \label{energy2}
\end{equation}%
For values of $\mu $ in the interval ($2.5<\mu <30$) one sees that the
values of $c_{1}$ and $c_{2}$ are close to the limiting values $c_{1}\simeq
c_{2}\simeq \frac{3}{16}$, while $c_{12}$ has the limiting value $%
c_{12}\simeq \frac{3}{4}$. (It is easy to obtain this limit by setting $%
\omega _{1}\sim \omega _{2}\sim \mu $ and $4\omega _{1}^{2}-\omega
_{2}^{2}\sim 4\omega _{2}^{2}-\omega _{1}^{2}\sim 3\mu ^{2}$ into the
formulas of (\ref{c1c2c12}) ). The limiting values for the critical energies
are $E_{c}^{(1)}=4(2-\sqrt{2})/3\simeq 0.78$ and $E_{c}^{(2)}=4(2+\sqrt{2}%
)/3=4.55$. The critical energy $E_{c}^{(1)}=0.78$, obtained for localized
excitations generated by $J_{2}=0$,$\ $agrees within twenty-five percent
with the numerical calculations of the next section, which determine $%
E_{c}\simeq 0.6$. For initial conditions in the 3-system, the energy leaking
out is compensated by a negative interaction energy of the 3-system with the
rest, such that the energy inside the 3-system is actually larger than the
total energy (this explains how we have overestimated the critical energy).
The reason for this better agreement is still that, by increasing the
subsystem size, the interaction energy with the immediate neighbors
(whatever its sign) becomes less important. Models with more oscillators in
the subgroup should furnish even better approximations, but they are harder
to work out analytically and the corresponding type($n$) initial conditions
are computationally more expensive to investigate.

\section{Numerical Results}

We present numerical results for the DNA oscillator chain, with initial
condition in two different types of oscillator groups. All of our numerical
integrations were performed with a tenth order symplectic
Runge-Kutta-Nystrom integrator\cite{Tsitouras}. The high-order integrator
can take very large steps, of about $0.6$ of the shortest linear period and
still conserves energy with a precision of $10^{-10\text{ \ }}$even after
integration times of $10^{10}.$

\subsection{Macroscopic quantities}

The dynamics of \ the full chain described by Lagrangian (\ref{NormalL}) is
described by the following Hamiltonian 
\begin{equation}
H=\dsum\limits_{i=-N/}^{N/2}\frac{1}{2}p_{i}^{2}+\frac{1}{2}(\xi _{i+1}-\xi
_{i})^{2}+\frac{1}{2}\mu ^{2}(\exp (-\xi _{i})-1)^{2}.  \label{Hamilto}
\end{equation}%
In numerical experiments the instantaneous values of the \emph{on-site}
oscillator energies $E_{i}$, $i=1,...,N,$ is usually calculated as 
\begin{equation}
E_{i}\equiv \frac{1}{2}p_{i}^{2}+\frac{1}{4}(\xi _{i+1}-\xi _{i})^{2}+\frac{1%
}{4}(\xi _{i+1}-\xi _{i})^{2}+\frac{1}{2}\mu ^{2}(\exp (-\xi _{i})-1)^{2},
\label{defEi}
\end{equation}%
where we include fifty percent of the interaction with the oscillator at
each side, such that the sum of the $E_{i}$ is the constant total energy.
Over short times the instantaneous and average values are nearly the same.
The information entropy is defined by 
\begin{equation}
S=-\sum_{i=1}^{N}e_{i}\ln e_{i},  \label{shanon}
\end{equation}%
where $e_{i}=E_{i}/\sum_{i}^{N}E_{i}$ are the normalized instantaneous
oscillator energies. In a typical situation where the total energy is
distributed among $r<N$ oscillators, $r$ of the $e_{i\text{ }}$are of order $%
1/r$ and the remaining are negligible, such that Eq. (\ref{shanon}) predicts 
$S$ $\simeq \ln (r).$ This motivates the definition of%
\begin{equation}
N_{osc}\equiv \exp (S),  \label{defNosc}
\end{equation}%
as the effective number of oscillators sharing the energy. It is also
convenient to define the normalized parameter 
\begin{equation}
n_{osc}\equiv N_{osc}/N.  \label{NoscN}
\end{equation}%
The normalized parameter $n_{osc}$ varies from $0$ to $1$, because the
entropy of Eq.(\ref{shanon}) is always less than $\ln (N).$ The
instantaneous value of $n_{osc}$ does not asymptote to one, due to
fluctuations. To calculate the effect of fluctuations we introduce a
deviation $\delta e_{i}$ from equipartition $e_{i}=\bar{e}+\delta e_{i}.$
Substituting this into (\ref{NoscN}), expanding the logarithm function as $%
\ln (1+\delta e_{i}/\bar{e})=\delta e_{i}/\bar{e}-(1/2)(\delta e_{i}/\bar{e}%
)^{2}$ and performing the summation over $i$ yields 
\begin{equation}
n_{osc}=\frac{1}{N}\exp \{-N\bar{e}\ln (\bar{e})-N\bar{(\delta e)^{2}}/(2%
\bar{e})\}=\exp \{-N\bar{(\delta e)^{2}}/(2\bar{e})\}.  \label{predictnef}
\end{equation}%
Taking $\bar{e}=1/N$ and making the assumption of normal statistics, that
for each normal mode $\bar{(\delta e)^{2}}=\bar{e}^{2}$ (which is true only
for the linearized lattice dynamics ), we see that $N$ cancels giving an
asymptotic value $n_{osc}=\exp (-0.5)=0.61$. This calculation shows that the
result does not depend on the number of oscillators if $N$ is large and also
shows why the value is different from unity. \ More accurate calculations
have been made, including the nonlinear terms in the oscillator calculation,
yielding\cite{Mirnov} 
\begin{equation}
n_{osc}=0.74  \label{neffvalues}
\end{equation}%
at equipartition of energy among the oscillators. These values have been
checked numerically, giving good agreement \cite{Mirnov}.

Numerical experiments show that for a randomly chosen localized initial
condition, the value of $n_{osc}$ usually starts to increase and reaches the
equipartition value $n_{osc}=0.61$ in a time of the order of $N$ , which is
the typical spread time. Our localization criterion is that a state is
localized when the value of $n_{osc}$ is significantly less than the
equipartition value $0.61$ for more than $50N$ periods $T_{f}$ of the
fastest linear mode ($T_{f}$ $=2\pi /\sqrt{\mu ^{2}+4}\sim 1$). For the
computationally accessible finite values of $N$ \ (of the order of $100$),
the smallest value of $n_{osc}$ is obtained for localization in a single
site, $n_{osc}=1/N$, which is an extreme value. \bigskip Given that $%
n_{osc}=0.61$ means equipartition, our practical criterion is $n_{osc}$ $<$ $%
\eta _{L}\equiv (20/N)$  for $t<50N$. \ With this criterion we give the state
some room to breathe, allowing the energy to spread over twenty oscillators
and then to shrink again to a smaller average value. This practical
criterion excludes either states localized in more than twenty oscillators
or states that would have sudden delocalization bursts, which was never
observed numerically. In the numerical calculations we use a logarithmic
scale for the increasing time, in the natural units of \ Lagrangian (\ref%
{NormalL}). The rapid fluctuations of the instantaneous values are smoothed
by taking the average of the last five instantaneous values of $n_{osc}$,
which are evaluated at a rate of 25 points per decade in time (at every
integer value of $25\ln (t)$).

Our numerical experiments integrate the dynamics of \ Lagrangian (\ref%
{NormalL}) for a chain of $N$ oscillators with periodic boundary conditions
and we shall use two types of initial conditions, defined as follows : (i)
Initial conditions produced by giving a nonzero position and momentum to a
single oscillator and a null value for the positions and momenta of all
other oscillators (the value of $N_{osc}$ at $t=0$ is 1). (ii) Symmetric
initial conditions produced by giving a nonzero value for three consecutive
oscillators with the symmetry $x_{-1}=x_{1}$ and $p_{-1}=p_{1}$ (the value
of $N_{osc}$ at $t=0$ is 3). \ For example we have used $\eta _{L}=0.2$ and
we started several (about 50) initial conditions of type (i) with a given
energy. For each initial condition we calculate $n_{osc}$ along the
numerical integration and we stop the integration at the first time that $%
n_{osc}$ becomes larger than $\eta _{L}=0.2$, defining a delocalization time
for that initial condition. The maximum value of the delocalization time ( $%
T_{\max }$ ) among the 50 initial conditions of the same energy is our
measure of localization. Typically, for a chain of $N=100$ oscillators, this
value is about $T_{\max }\simeq 100=N$ for subcritical energies, than there
is a rapid transition where this value climbs to above $T_{\max }=5000$. In
practice, it is necessary to stop the numerical integration in the
supercritical region whenever $T_{\max }$ reaches a maximum value, and we
have used $T_{\max }=53N$ as a good computationally accessible large number (%
$53N=5300$\ if $N=100$). We experienced with a much higher threshold for $%
T_{\max }$ of about \ $1000N$\ and obtained virtually the same type of
transition, but the numerical experiment becomes very time consuming. The
question if this localization time is either infinite, exponentially long,
or simply a very large value is not addressed in the present work. We have
also varied the threshold value of $n_{osc}$ among the values $\eta
_{L}=0.15 $, $\eta _{L}=0.2$ and $\eta _{L}=0.25$ and we obtained the same
transition line. We used for $N$ the three values $N=100$, $N=200$ and $%
N=500 $ and obtained virtually the same transition lines for $\mu >3$. A
comprehensive statistical analysis has not been performed due to the very
long times for some runs. Spot checks for a few cases indicate that the
spread from varying $N$ and $\eta _{L}$ is less than some few percent if $%
\mu >3$ and $\eta _{L}<(25/N).$ For the region $2.5<\mu <3$ there can be
significant changes in the critical energies determined by the above
procedure. This is because close to the critical value $\mu =2.25$ the
localization length becomes long, and in a lattice with a small $N$ this
localization is confused with equipartition by our criterion. Interesting to
recall that biology chose the safe value of $\mu =6.3$ possibly for the same
reason.

In Fig. 2 we plot the value of $T_{\max }<53N$ among 49 initial conditions
of type (i) as a function of the energy for $\mu =6.3$. Notice the
pronounced jump in $T_{\max }$ which is a signature of localization. We
define the critical energy by the inflection point of the $T_{\max }$ curve,
which from Fig. 2 is about $E\simeq 2.3$. This same discontinuous behavior
of $T_{\max }$ is observed in the numerical calculations for $2.5<\mu <30$,
and in Fig. 3 we plot the critical energy determined by the inflection point
of $T_{\max }$ and the theoretical predictions for the 1-system, Eq.(\ref%
{Ecritical1}), versus $\mu $ . The numerically determined critical energy is
about twice the predicted for the simple 1-system by perturbation theory.
This effect is due to the fact that for type (i) initial conditions a
substantial part of the energy leaks to the immediate neighbors even when
there is localization, such that the total energy of the system at
localization is significantly larger than the energy of the 1-system.

In Fig. 4 we plot $T_{\max }<53N$ among 49 type (ii) initial conditions, as
a function of the energy for $\mu =3.0$, and $N=100$, illustrating the same
jump that is our signature of localization. The critical energy predicted by
the inflection point of Fig. 4 is $E=0.75$. In Fig. 5 we plot the critical
energy determined by the inflection point of $T_{\max }$ and the theoretical
predictions of the 3-system versus $\mu $. The theoretical prediction for
the 3-system agrees with the numerical results within twenty five percent.
The approximation is better than in the case of the 1-system because less
energy leaks out of the 3-system. For type(ii) initial conditions there is
an interaction term in the total energy that increases the energy inside the
3-system above the total energy, but the predicted energy is now only twenty
five percent wrong.

In Fig. 6 we plot the modulus of the complex Fourier transform of the
coordinate of \ the central oscillator for an initial condition of type (i)
of a lattice with $\mu =6.3$ , $N=100$ and a subcritical energy $E=0.1$.
Notice that the Fourier transform develops nonzero components inside the
conduction band $6.3<\omega <6.61$, as illustrated in the insert to Fig. 6.

In Fig. 7 we plot the Fourier transform of the coordinate of \ the central
oscillator for a localized initial condition of type (i) in a lattice with $%
\mu =6.3$, $N=100$ and a supercritical energy $E=3.0$, which has a primary
peak at $\omega =6.0<\mu $ and goes to zero already at $\omega =6.25<\mu $,
in accordance with the (CC) conjecture.

Last, in Fig. 8 we plot the surface of section of the 3-system with $\mu
=6.3 $ at the supercritical energy $E=3$, showing very little stochasticity,
to illustrate that localization has nothing to do with stochasticity within
the subgroup, as discussed in Ref.\cite{FW98}.

\section{\protect\bigskip Discussions and conclusion}

At a supercritical energy, by searching among 49 initial conditions of type
(i), for example, we have found several initial conditions that stay
localized for more than $10^{5}$ \ natural units. Using a numerical search
that varies the initial condition in the neighborhood of an original
localized condition \cite{simple}, in a way that maximizes the localization
time, we could easily find other initial conditions that stays localized for
a much larger time, of the order of $10^{7}$. These refined initial
conditions become restricted to narrow domains, and we believe that the
study of timescales for a localized excitation in a chain with a finite $N$
should start from here in future work, for example to test if one can
increase this time arbitrarily.

At a finite $N$, if the energy of a type (ii) initial condition does not
localize in the original 3-system, it will leak out to the other $(N/3)$
3-systems of the chain. A simple condition for these other 3-systems to be \
"sufficiently linear" is then that the total energy be less than $NE_{c}/3$
\ (such that the other 3-systems display a quasi-linear motion). This
intensive condition, $E<$ $NE_{c}/3$, is important to remember in numerical
experiments with a finite lattice. For example for a chain of $N=100$
oscillators, this means $E<26.4$ and in our numerical experiments we have
always stayed well below this energy.

The critical value of $\mu $ for localization ($\mu =2.25$ ) is in agreement
with the numerical calculations, as we never found localization below $\mu
=2.5$. In the region $2.5<\mu <3.0$, the numerical results indicate that the
localization length is very large, which requires numerical experiments with
large values of $N$. The value $\mu =6.3$ estimated from the biological
measurements is far from the critical and in a region where localization
length is small, such that we predict a robust localization from the above
DNA model. The threshold energy for localization at $\mu =6.3$ is $2.2$
units or $4.4\times 10^{-3}$ $eV$ ( $0.17k_{B}T$ at room temperature). This
means that localization is possible at room temperature, as predicted by our
model. Last, the localization time found numerically is greater than $10^{5}$
time units or $10^{-7}$ seconds, enough for the biochemical mechanisms to
operate. Such localization can be related to the bubbles in DNA and it would
be an auxiliary mechanism in the transcription process.

\section{Acknowledgements}

We thank Allan Lichtenberg for discussions on breather research. We
acknowledge a Fapesp grant which supported A. Ponno in Brazil, where this
work was completed. A. Ponno acknowledges GNFM for the payment of travel
expenses. \ J. De Luca and E. Drigo Filho acknowledge the partial support of
a CNPQ scholarship.

\section{\protect\bigskip Figure Captions}

Fig. 1 Frequency of the nonlinear 1-system divided by $\mu $, $(\omega /\mu
) $, plotted as a function of the energy for $\mu =2$ ( dashed line), $\mu
=2.25$ (solid line) and $\mu =2.5$ (dotted line ). The horizontal solid line
is $\omega /\mu $ $=1.$ Notice that at $\mu =2.25$ the frequency line is
only tangent to the critical line.\bigskip\ Arbitrary units.

Fig. 2 $T_{\max }$ as a function of the energy for type (i) initial
conditions at $N=100$ and $\mu =6.3$. Arbitrary units. The squares represent
numerical calculations and the solid line is spline interpolation.

\bigskip

Fig. 3 Numerically calculated critical energy for type (i) initial
conditions with $N=100$ (triangles), $N=200$ (stars) and $N=500$ (circles)
as a function of the energy. Also plotted is the critical energy \ predicted
by the 1-system (squares ). Arbitrary units.

\bigskip

Fig 4. $T_{\max }$ as a function of the energy for type (ii) initial
conditions at $N=100$ and $\mu =3.0$(triangles). Arbitrary units. The
triangles represent the numerical calculations and the solid line is spline
interpolation.

\bigskip

Fig. 5 \ Numerically calculated critical energy for type (ii) initial
conditions with $N=100$ as a function of $\mu $ (triangles) and critical
energy predicted by the 3-system $E_{c}^{(1)}$(circles), as a function of $%
\mu $. Arbitrary units.

\bigskip

\bigskip Fig. 6 Fourier transform of an initial condition of type (i) with $%
E=0.1$ (subcritical) \ for a lattice with $\mu =6.3$ and $N=100$. Plotted is
the modulus $F(\omega )$ of the complex Fourier transform. The insert
magnifies the region near $\omega =6.3$ to display that $F(\omega )$ is not
zero inside the conduction band.

Fig.7 \ Fourier transform of an initial condition of type (i) with $E=3.0$ \
for a lattice with $\mu =6.3$ and $N=100$. Plotted is the modulus $F(\omega
) $ of the complex Fourier transform. The insert magnifies the region near $%
\omega =6.0$ to display the peak of $F(\omega )$ at $\omega =6.0<\mu .$
Notice that $F(\omega )$ vanishes above $\omega =6.25<\mu .$

\bigskip

Fig. 8 Surface of section of the symmetric 3-system at $\mu =6.3$ and $E=3.0$%
, showing little stochasticity. Arbitrary units.

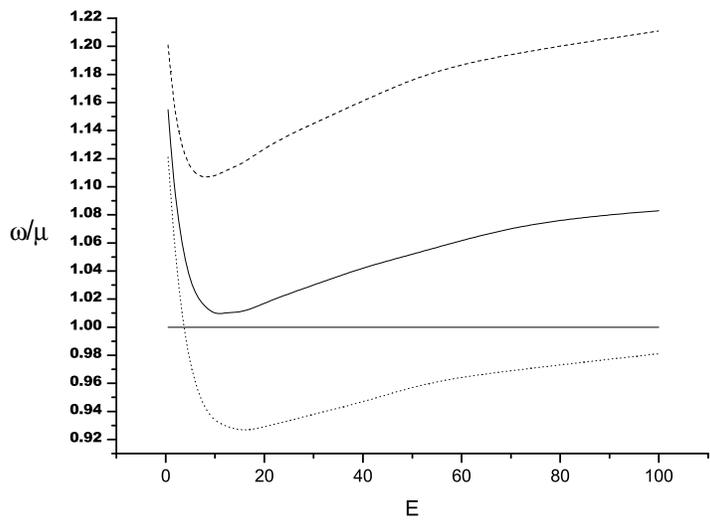
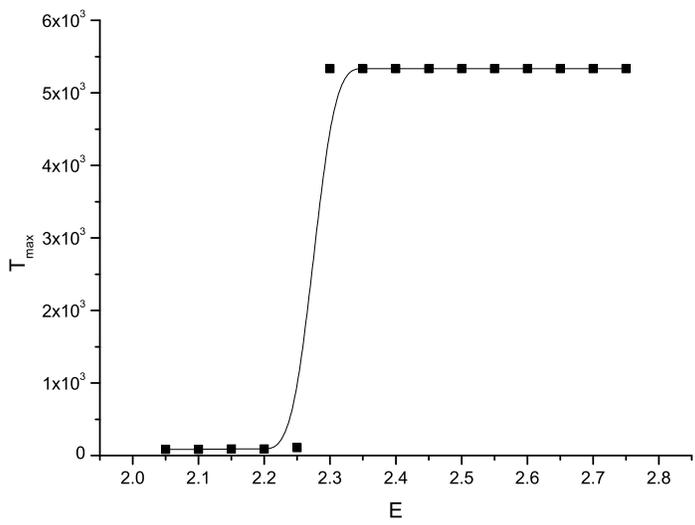
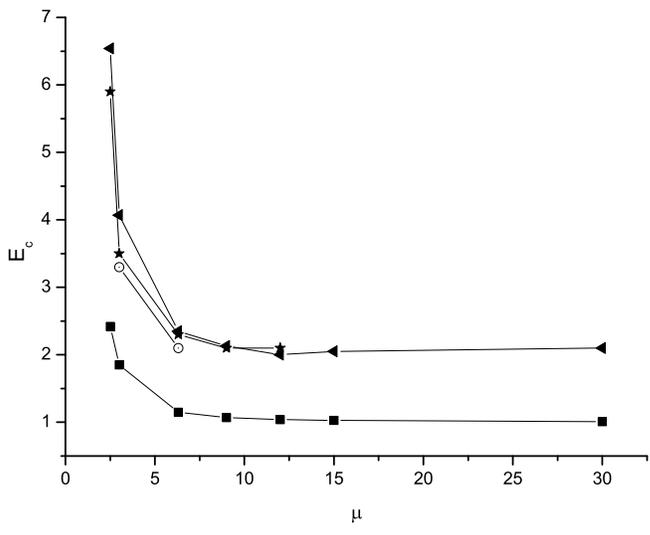
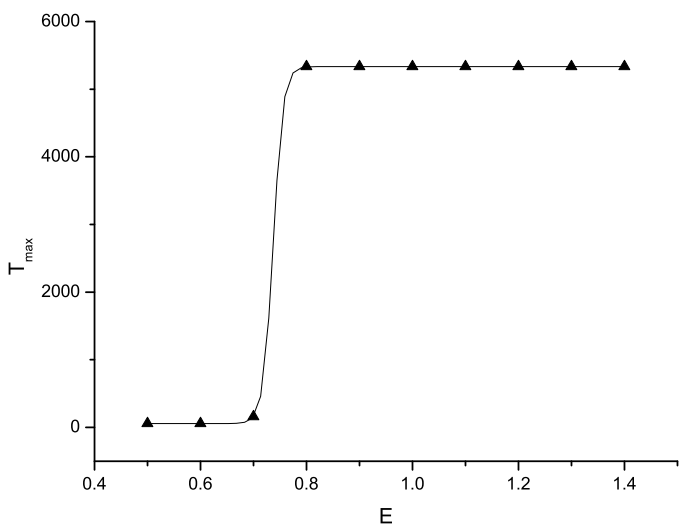
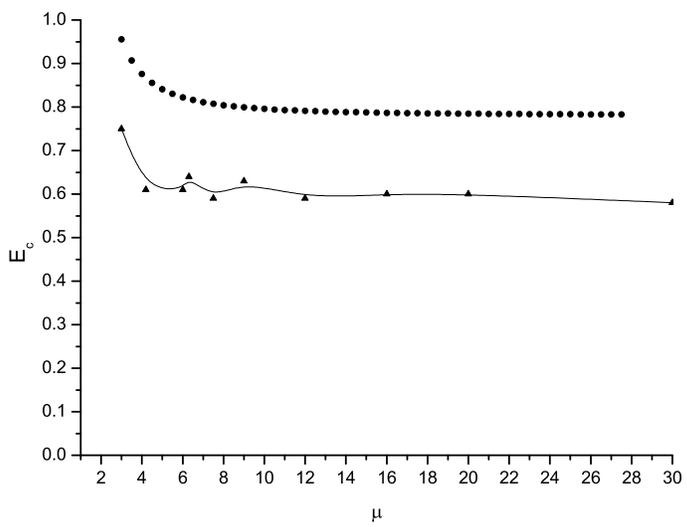
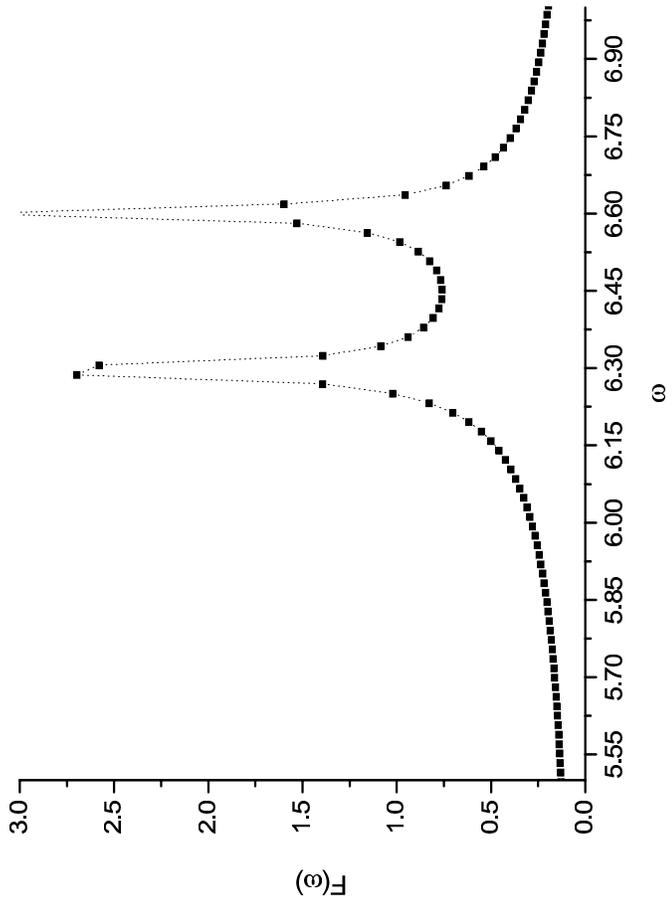
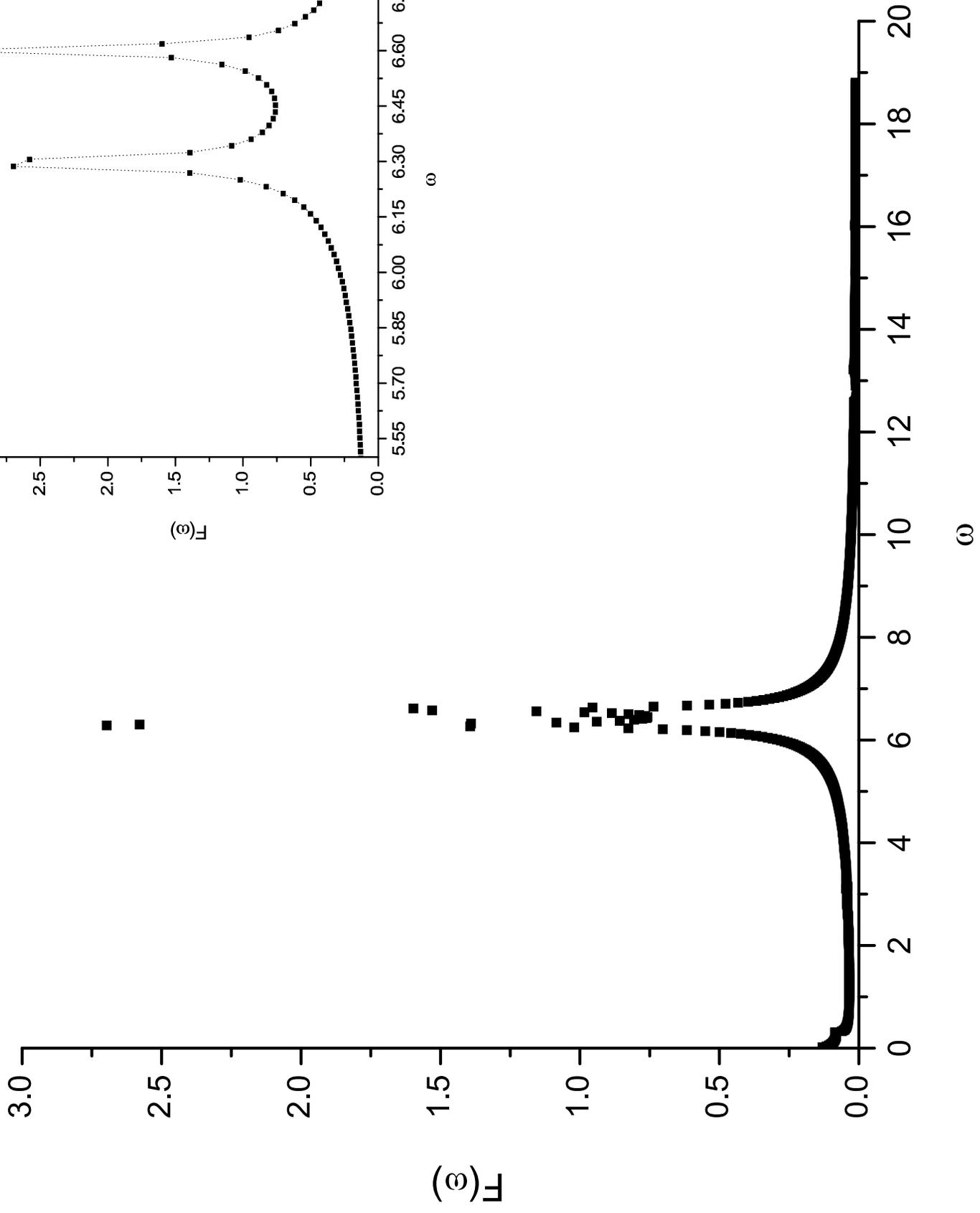
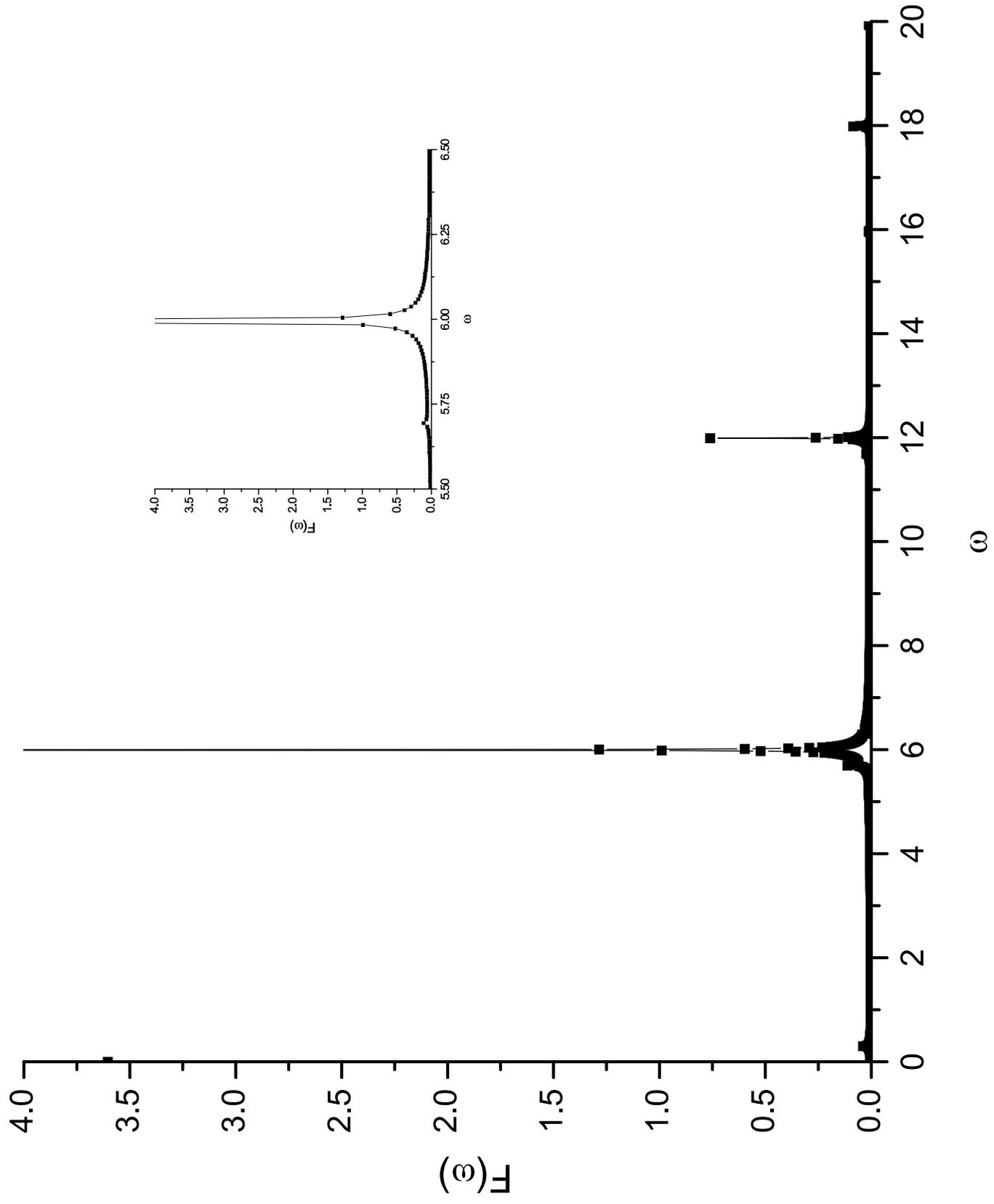
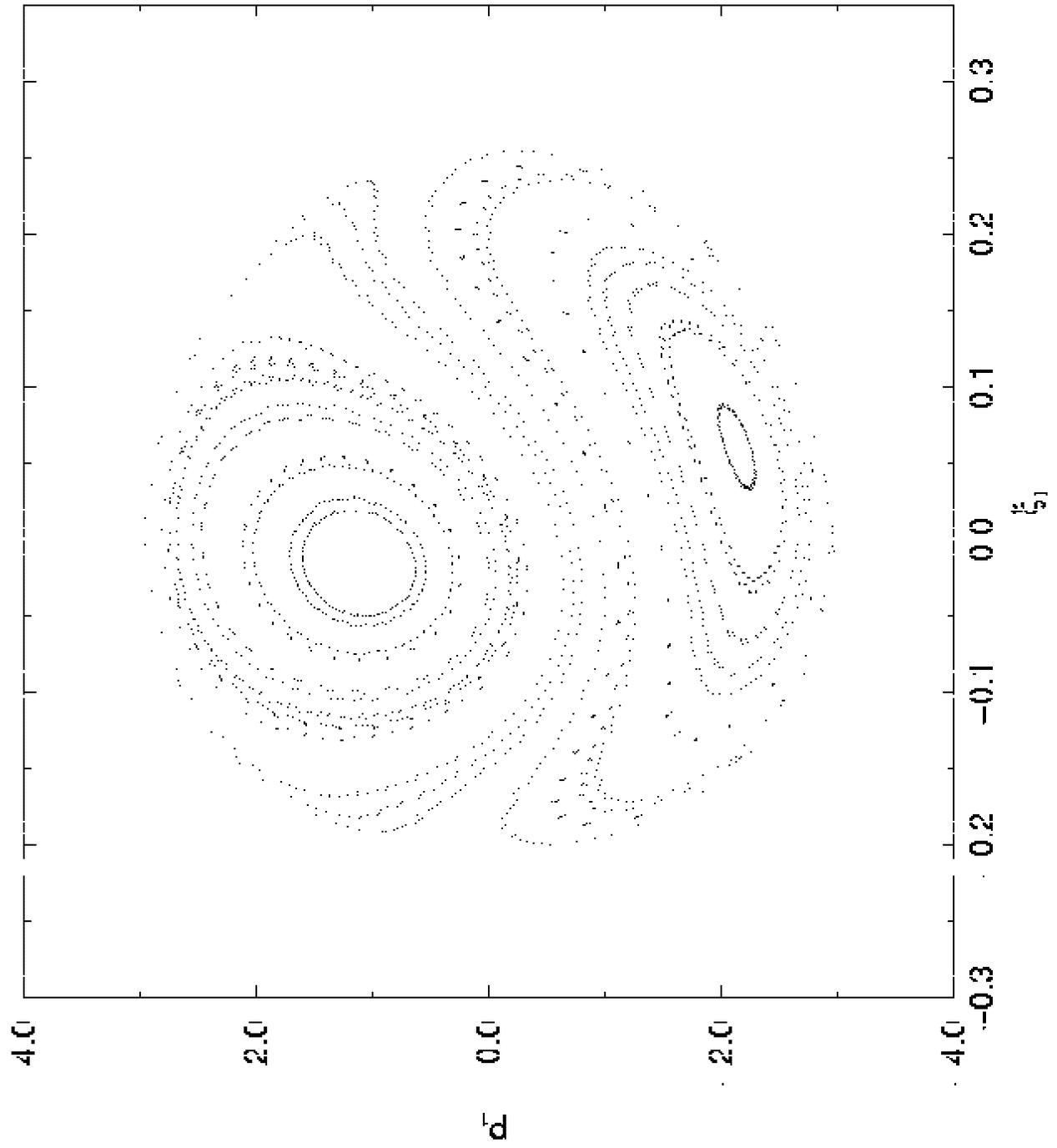
\end{document}